\documentclass{emulateapj}
\usepackage{natbib}
\usepackage{amssymb}
\usepackage{amsmath}
\usepackage{graphicx}

\def \Ni {$^{56}$Ni}
\def \tN {{t_{\rm Ni}} }
\def \Lb {L_{\rm bol} }
\def \Le {\Lambda_e }
\def \Q {Q_{\rm Ni} }
\def \eN {\eta_{\rm Ni} }
\def \dm {\Delta M_{25-75} }

\def \Et {E_{r,0}t_0 }
\def \Eexp {E_{\rm exp}}
\def \Eenv {E_{\rm env}}
\def \Mej{M_{\rm ej}}
\def \vej{v_{\rm ej}}
\def \venv{v_{\rm env}}
\def \Menv{M_{\rm env}}
\def \Renv{\overline{R}_{\rm env}}

\begin{document}
\title{The importance of \Ni\ in shaping the light curves of type II supernovae} 
\author{Ehud Nakar\altaffilmark{1}, Dovi Poznanski\altaffilmark{1} and Boaz Katz\altaffilmark{2}}
\altaffiltext{1}{The Raymond and Beverly Sackler School of Physics and Astronomy, Tel Aviv University, Tel Aviv 69978, Israel}
\altaffiltext{2}{Department of Particle Physics and Astrophysics, Weizmann Institute of Science, Rehovot 76100, Israel}

\email{}

\begin{abstract}
	What intrinsic properties shape the light curves of Type II supernovae (SNe)? To address this question we derive observational measures that are robust (i.e., insensitive to detailed radiative transfer) and constrain the contribution from \Ni, as well as a combination of the envelope mass, progenitor radius, and explosion energy. By applying our methods to a sample of type II SNe from the literature we find that \Ni\ contribution is often significant.  In our sample its contribution to the time weighted integrated luminosity during the photospheric phase ranges between 8\% and 72\% with a typical value of 30\%.  We find that the \Ni\ relative contribution is anti-correlated with the luminosity decline rate. When added to other clues, this in turn suggests that the flat plateaus often observed in type II SNe are not a generic feature of the cooling envelope emission, and that without \Ni\ many of the SNe that are classified as II-P would have shown a decline rate that is steeper by up to 1 mag/100 d. Nevertheless, we find that the cooling envelope emission, and not \Ni\ contribution, is the main driver behind the observed range of decline rates. Furthermore, contrary to previous suggestions, our findings indicate that fast decline rates are not driven by lower envelope masses. We therefore suggest that the difference in observed decline rates is mainly a result of different density profiles of the progenitors.
\end{abstract}

\section{Introduction}
Observations and modeling of supernova (SN) light  indicate that the two most common power sources of the emission we observe are the radioactive decay of \Ni\ and the internal energy deposited in the envelope by the shock that unbinds the star \citep[][and references therein]{Woosley02}. 
The emission powered by the latter is known as the cooling envelope emission and it dominates the photospheric phase in most type II SNe \citep{Falk77} that mark the explosion of a red supergiant (RSG)\footnote{ In the majority of type II SNe the photospheric phase seems to be dominated by cooling envelope emission. There are however exceptions. For example in type II-b and 1987A-like SNe the photospheric phase is dominated by radioactive decay of \Ni, while in type II-n SNe interaction is likely to contribute significantly. Here we consider only SNe of types II-P and II-L referring to them simply as type II SNe.} \citep[][ and references therein]{Smartt09B,Smartt09A}. These are sub-classified into two sub-types based on the luminosity decline rate during that phase. If the photospheric phase shows no (or at most a moderate) decline the SN is classified as II-P while if the decline is faster it is classified as II-L \citep{Barbon79}. This classification is not well defined and different authors use different decline rates in different bands in order to separate the two classes \cite[e.g., ][]{Patat94,Arcavi12,Faran14,Faran14a}. Earlier studies suggested that the decline rate separates type II SNe into two distinctive populations \cite[e.g.,][]{Patat94,Arcavi12}. However, recent studies (\citealt{Anderson14,Faran14,Sanders15}, see however \citealt{Poznanski15}) suggest that there is a continuous distribution of decline rates, thus challenging the usefulness of this classification. In this paper we do not attempt to separate SNe into these two categories, using instead the measured decline rate to characterize  each SN.

Interestingly, there is no clear theoretical explanation for the origin of the difference in the decline rates (be it continuous or not), or to the correlation of the decline with other SNe properties such as the peak brightness \citep{Anderson14,Faran14} or spectral features \citep{Schlegel90,Faran14,Gutirrez14}. The most common, yet unconfirmed, suggestion is that steeper decline rates are generated by smaller envelope masses \cite[e.g.,][]{Barbon79,Swartz91,Blinnikov93}.

Numerical modeling of RSG explosions that include \Ni\ suggest that although cooling envelope emission dominates the photospheric phase, the power deposited by \Ni\ may have an observable effect also during that phase \citep[e.g.,][]{Falk77,Young04,Utrobin07,Utrobin09,Kasen09,Bersten11, Dessart11,Dessart13}. The main effect of \Ni\ on the light curve is via the additional radiated energy, which results in a brighter emission. A second order \Ni\ effect is the increase in ionization fraction (and thus the opacity) of the envelope, thereby delaying the release of internal energy deposited both by the shock and by the \Ni\ itself. For any realistic distribution of \Ni, its contribution to the observed emission increases with time during most of the photospheric phase (this is true even for a uniform \Ni\ mixing through the entire star; e.g., \citealt{PiroNakar13}). This may make the final stages of the photospheric phase brighter, and possibly more extended.

The purpose of this paper is to use observations of type II SNe to study the impact of \Ni\ decay on the photospheric phase emission. In particular we are interested in a quantitative measurement of the \Ni\ importance in observed SNe and in separating \Ni\ contribution from that of the cooling envelope emission, which is powered by shock deposited energy. We pay special attention to the effect \Ni\ has on the light curve decline rate.

The paper structure is as follows.  In \S\ref{sec:theory} we use exact energy conservation arguments to
derive  a robust  measurement of  the contribution  of \Ni\  during the  photospheric phase,  using the
time-weighted integrated  bolometric light  curve. We  also derive  a measure  of the  cooling envelope
emission alone (i.e.,  the one that would have been  observed if there were no \Ni),  which is directly
related to its light curve shape. In  section \S\ref{sec:NiContribution} we apply these measurements to
the bolometric light curves of a sample of SNe that were compiled from the literature and are described
in \S\ref{sec:sample}. We  show that \Ni\ is important  in many SNe, and constrain the effect that \Ni\ has on the light curve shape. In \S\ref{sec:correlations} we test for correlations between the observables we introduce here and other observables we obtain for our sample.  Finally,  in
\S\ref{sec:SN_properties} we use the time-weighted light curve  to obtain a new global measurement that
constrains the  progenitor radius, envelope  mass, and explosion  energy, by rigorously  subtracting the
\Ni\ contribution. We show that, contrary to previous suggestions, a faster decline rate is most likely
not  a result  of a lower  envelope mass. We  summarize our
results in \S\ref{sec:Summary}.

\section{Separating the \Ni\ contribution from the cooling envelope emission}\label{sec:theory} 

We are interested in a well-defined observationally-based measure that can separate the contributions of cooling envelope and \Ni\ decay to the photospheric emission for a given bolometric light curve $\Lb(t)$. In type II SNe there is a sharp drop in the light curve once the photosphere ends crossing the envelope, allowing a clear distinction between the photospheric phase and the \Ni\ tail. We denote this time as $\tN$, and in well observed SNe it can be determined to within 10\,d accuracy or better. At $t>\tN$ the diffusion time through the envelope is much shorter than $t$ and therefore $\Lb(t>\tN)=\Q(t>\tN)$, where
\begin{equation}\label{eq:QNi}
\Q(t)=\frac{M_{Ni}}{M_{\odot}}(6.45 e^{-t_d/8.8}+1.45e^{-t_d/111.3})\times 10^{43}{\rm erg\,s^{-1}}
\end{equation}
is the instantaneous injection of energy into the ejecta due to the \Ni\ decay chain. $t_d$ is time since the explosion in days. Escape of gamma-rays generated by the decay of \Ni\ is minor in type II SNe for some time after $\tN$, as evident from the \Ni\ tail observed decay rates, which are typically consistent with Eq. \ref{eq:QNi}. Therefore, observations of the \Ni\ tail provide  a rather accurate estimate of the total \Ni\ mass in the ejecta\footnote{Some gamma-ray escape may take place if \Ni\ is well mixed throughout the envelope, which can lead to a slight underestimate of $M_{Ni}$  \citep[e.g.,][]{Morozova15}.}, $M_{Ni}$. Once $M_{Ni}$ is found, the \Ni\ injection rate of energy $\Q(t)$ at any given time is determined using \eqref{eq:QNi} (up to late times where gamma-ray escape becomes significant). Now we are faced with the following question -- at early times $t<\tN$, how much of $\Lb(t)$ was due to the cooling envelope and how much was due to $\Q(t'<t)$?

There are a few challenges that do not allow a simple separation of the two contributions:
\begin{itemize}
\item Energy which is deposited at some time $t'$ is released at an unknown later time $t$.
\item The opacity depends on the temperature and composition and is thus affected in a non linear way by the contribution of the two energy sources.
\item There are continuous adiabatic losses so that the emitted energy is smaller than the injected energy.
\end{itemize}

Let us imagine first that the ejecta were not expanding, so that there were no adiabatic losses. In such a case, we would know that the total emitted energy $\int \Lb\,dt$ has to be the sum of the shock deposited energy and the integrated injected energy $\int\Q(t)\,dt$. Since we know $\Lb(t)$ and $\Q(t)$, we would be able to determine each of the contributions. Note that this separation would be based on total energy conservation and would thus not depend on the difference between deposited and released times nor on the effect of \Ni\ decay on the opacities. If we were to apply the same energy conservation argument to an expanding ejecta it would fail due to the significant and unknown adiabatic losses. 

A way to follow the conservation of total energy in an expanding ejecta was recently realized in the context of type Ia SNe \citep{Katz13} and we apply the same method here. The idea is that for non-relativistic homologous expansion, where the energy is dominated by radiation, adiabatic losses do not affect the product $E_rt$, where $E_r$ is the energy in radiation that is trapped in the ejecta and $t$ is the time since explosion. Thus we can separate the contributions by using the time-weighted emission $\int t\,\Lb\,dt $. 

For clarity we repeat here these arguments and extend them to include the cooling envelope emission which is negligible in SNe Ia. During the homologous phase, which starts in the case of RSG progenitors about a day after the explosion, the derivative of the internal energy (which is dominated by radiation) in the outflow satisfies:
\begin{equation}\label{eq:dE_dt}
	\frac{dE_r(t)}{dt}=-\frac{E_r(t)}{t}+\Q(t)-\Lb (t) .
\end{equation} 
The term $-E_r/t$ is the total rate of adiabatic losses. After rearranging the equation, multiplying both sides by $t$ and
integrating over $t$ from $0$ to $\tN$ one obtains: 
\begin{equation}\label{eq:E0t01}
	\int^{\tN}_0 t\,\Lb\,dt=\Et+\int^{\tN}_0t\,\Q\,dt,
\end{equation}
where $t_0$ is the time at which the homologues phase begins, $E_{r,0}=E_r(t_0)$ and we used the fact that the contribution at $t<t_0$ to the integral on the r.h.s of the equation is negligible. We also used the fact that at late enough time $t \geq \tN$ the diffusion time through the envelope is much shorter than $t$ and therefore $E_rt \rightarrow 0$. On the left hand side of Eq. \eqref{eq:E0t01}, we have the total (time weighted) energy released. On the right side we have the total (time weighted) energy deposited, where we separated contributions from the \Ni\ decay chain $\int^{\tN}_0 t\,\Q\,dt$ and from the shock deposited energy $\Et$ (just like in the hypothetic static case, but with time weights). The measurable parameter $\Et$ is of interest by itself as it is related directly to the progenitor structure and mass and to the explosion energy. We explore this relation in \citet{Shussman16}. We use their results here (section  \ref{sec:SN_properties}) to study the sample of SNe that we compile in section  \ref{sec:sample}. We use here the notation of \citet{Shussman16} for $\Et$: $ET \equiv \Et$ 

Next, we define $L_e$ as the luminosity that would have been seen if there were no \Ni\ present in the ejecta. Since without \Ni\ $\Lb=L_e$, Eq. (\ref{eq:E0t01}) implies $\int^\infty_0t\,L_e\,dt =\Et$. This in turn implies that when \Ni\ does exist (but $L_e$ is still the hypothetical luminosity when it does not):
\begin{equation}\label{eq:ET2}
 ET=\int^\infty_0 t\,L_e\,dt =	\int^{\tN}_0 t\,\Lb\,dt-\int^{\tN}_0t\,\Q\,dt.
\end{equation}

An observable measure of the \Ni\ contribution to the emission during the photospheric phase ($t<\tN$) can be defined as the ratio of the (time weighted) energy deposited by \Ni\ decay to the shock deposited energy (multiplied by $t_0$):
\begin{equation}\label{eq:etaNi}
\eN \equiv \frac{\int^{\tN}_0 t\,\Q\,dt}{\Et}=\frac{\int^{\tN}_0 t\,\Q\,dt}{\int^{\infty}_0 t\,L_e\,dt}= \frac{\int^{\tN}_0 t\,\Q\,dt}{\int^{\tN}_0 t\,(\Lb-\Q)\,dt}.
\end{equation} 
$\eN \ll 1$ implies that shock deposited energy is the main power source and cooling envelope emission dominates during the entire photospheric phase. $\eN \gg 1$ indicates that \Ni\ decay dominates the energy output during most of the photospheric phase. When $\eN \approx 1$ the two power sources have a comparable contribution where cooling envelope dominates at early time and \Ni\ is more dominant near the end of the photospheric phase.

Finally, we define another dimensionless observable:
\begin{equation}\label{eq:Lambda_e}
	\Le \equiv \frac{L_{e,25} \cdot(80\,d)^2}{\int^\infty_0 t\,L_e\,dt} = \frac{L_{25} \cdot(80\,d)^2}{\int^{\tN}_0 t\,(\Lb-\Q)\,dt},
\end{equation} 
where $L_{e,25}$ and $L_{25}$ are the hypothetical $L_e$ and the observed $\Lb$ on day 25, respectively. As the \Ni\ effect on $\Lb$ is expected to be negligible on day 25 we assume that $L_{e,25}=L_{25}$. The rational behind this definition is that $\Le$ is an observable that depends purely on $L_e$. Being the ratio between $L_{e,25}$ and the time-weighted integrated $L_e$ it measures a combination of the decline rate and duration of the light curve had \Ni\ been absent. In section \ref{sec:NiContribution} we discuss its interpretation in more detail. The constant $(80\,d)^2$ is inserted to make $\Le$ a dimensionless parameter of order unity, where the specific choice of $80\,d$ is explained in section \ref{sec:NiContribution}.

We therefore have defined three observables, $\eN$, $\Le$, and ET, that will allow us to separate and evaluate the relative contributions of the cooling envelope and \Ni\ to a given light curve.

\begin{deluxetable*}{p{1cm}cccccccccccc}
\tabletypesize{\scriptsize}
\tablecolumns{13} 
\tablewidth{0pt}
\setlength{\tabcolsep}{0.0in}
 \tablecaption{The bolometric sample - observed SNe with bolometric or opt/IR luminosity}
 \tablehead{
 SN & $\eN$ & $2.5\log(\Le)$ & $\dm$ &$ET$ & $L_{25}$ & $L_{50}$ & $L_{75}$ &$\tN$ & $M_{Ni}$$^\dagger$& L type$^{\dagger\dagger}$ &$v_{50}$$^{\dagger\dagger\dagger}$& Ref.\\
& & & &[$3\cdot 10^{55}{\rm erg\,s}$] &\multicolumn{3}{c}{[$10^{41}{\rm erg/s}$]} & [d] &[$0.01M_\odot$] &  & [${\rm km/s}$]}
\startdata
1999em 		& 0.54 & 0.55& 0.17 & 1.34 & 14	    & 12 	& 12   & 136 & 4.7 & bol 	&3280 &  1\\
1999gi 		& 0.63 & 0.66& 0.33 & 1.37 & 16	    & 13 	& 12   & 130 & 5.8 & bol 	&3700 & 1,2\\
2003hn 		& 0.31 & 1.3 & 0.85 & 1.16 & 23 	& 13 	& 11   & 110 & 3.2 & bol 	&& 1\\
2004et 		& 0.43 & 0.77& 0.53 & 1.83 & 23 	& 16 	& 14   & 136 & 5.0   & UBVRIJHK &4020 & 3\\
 			&(0.32)&(0.85)&(0.70)&(1.23)& (17)	& (10) 	& (8.8)&(136)& (2.5) &(UBVRI) & \\
2005cs$^a$ 	& 0.26 & 0.08& -0.09 & 0.28& 1.9 	& 2.0 	& 2.1  & 130 & 0.5 & bol 	&1980 &   4\\
 			&(0.15)&(0.11)&(0.01)&(0.17)&(1.2)  &(1.1) 	&(1.2) &(130)&(0.17) & (UBVRI)& \\
2007od$^a$ 	& 0.12 & 0.97& 0.87 & 2.12 & 32 	& 22 	& 15   & 112 & 2.0 & uvoir &3130 & 5\\
2009N 		& 0.61 & 0.67& -0.04 & 0.40 & 4.6 	& 5.1 	& 4.8  & 112 & 2.0 & bol 	&2580 &  6\\
 			&(0.45)&(0.68)&(0.09)&(0.21)&(2.5)  &(2.5)  &(2.3) & (112)&(0.8)& (BVRI) && \\ 
2009ib      & 2.6  & 1.1 & 0.1  & 0.29 & 5.2    & 4.8   & 4.7  & 142 & 4.6 & uvoir  &3090 &   7\\
2009md 		& 0.28 & 0.59& 0.26 & 0.23 & 2.4 	& 2.2 	& 1.9  & 121 & 0.5 & UBVRIJHK & 2000$^c$ & 8\\
2012A 		& 0.29 & 0.91& 0.56 & 0.42 & 6.1 	& 4.6 	& 3.6  & 117 & 0.9 & uvoir 	& 2840$^c$& 9\\
2012aw$^b$ 	& 0.49 & 0.56& 0.31 & 1.52 & 16 	& 13 	& 12   & 135 & 4.9 & uvoir	&3890 & 10\\
2012ec 	    & 0.57 & 0.88& 0.18& 0.73  & 10 	& 9.3 	& 8.8  & 112 & 3.5 & UBVRIJHK &3370 & 11\\ 
2013by 		& 0.2 & 1.2 & 1.19 & 1.66 & 33     & 25	& 11   & 104 & 2.9 & bol 	&& 12, 13
\enddata
 \tablenotetext{}{1. \cite{Bersten09} 2. \cite{Leonard02} 3. \cite{Maguire10} 4. \cite{Pastorello09} 5. \cite{Inserra11} 6. \cite{Takats14} 7. \cite{Takats15}
8. \cite{Fraser11} 9. \cite{Tomasella13} 10. \cite{DallOra14} 11. \cite{Barbarino15}  12. \cite{Valenti15} 13. This paper} 
\tablenotetext{$^\dagger$}{Derived based on comparison of Eq. \ref{eq:QNi} to the observed luminosity, without additional bolometric corrections (see text).} 
\tablenotetext{$^{\dagger\dagger}$}{{\it `bol'} light curves include bolometric correction factor or calculated based on blackbody fits. All other types are the integrated luminosity in the observed bands without additional bolometric corrections. }
\tablenotetext{$^{\dagger\dagger\dagger}$}{Typical velocity error is of order 200 km/s.}
\tablenotetext{a}{Inaccurate $M_{Ni}$. The tail does not follow the decay rate predicted by \Ni\ decay, making the estimate of the \Ni\ mass less accurate. In SN 2007od, where dust may affect the tail emission, we use the \Ni\ mass estimates from \cite{Inserra11}.}
\tablenotetext{b}{Sparse data. The transition time to the \Ni\ tail is not observed. It is between 130 d (during the sharp fall near the end of the photospheric phase) and 286 d (first data point in the tail). Here we take $\tN=135$ d.}
\tablenotetext{c}{$v_{50}$ is taken from the literature.}
\label{table:BolSample}
\end{deluxetable*}

\begin{deluxetable*}{p{1cm}cccccccccccc}
\tabletypesize{\scriptsize}
\tablecolumns{11} 
\tablewidth{0pt}
\setlength{\tabcolsep}{0.0in}
 \tablecaption{The optical sample - observed SNe with UV/opt or optical only luminosity}
 \tablehead{
 SN & $\eN$ & $2.5\log(\Le)$ & $\dm$ & $L_{25}$ & $L_{50}$ & $L_{75}$ &$~~\tN~~$ & $M_{Ni,opt}$$^\dagger$& L type$^{\dagger\dagger}$ &  Ref.\\
& & & &\multicolumn{3}{c}{[$10^{41}{\rm erg/s}$]} & [d] &[$0.01M_\odot$] &  & [${\rm km/s}$]}
\startdata
1992H$^a$ 	& 0.71 & 1.1 & 0.70  & 17	& 11 	& 9.0  & 142 & 4.2 & BVR  &    1\\
1995ad$^b$ 	& 0.4  & 1.4 & 0.83  & 9.1	& 6.2 	& 4.2  & 98  & 1.5 & UBVRI 	&  2\\
2001dc 		& 0.43 & 0.82& 0.35  & 0.90 & 0.73 	& 0.65 & 120 & 0.22 & BVRI 	& 3\\
2003Z$^{c}$ & 0.27 & 0.57& 0.24  & 0.95 & 0.84 	& 0.76 & 130 & 0.21 & BVRI 	& 4\\
2004A 		& 0.63 & 0.68& 0.06  & 4.4 	& 4.3 	& 4.2  & 120 & 1.75 & BVRI 	&  5\\
2008in 		& 0.39 & 0.96& 0.53  & 2.8 	& 2.1 	& 1.7  & 115 & 0.55 & BVRI 	& 6\\
2009bw 		& 0.17 & 0.64& 0.77  & 13 	& 7.8 	& 6.6  & 140 & 1.2 & UBVRI 	& 7\\ 
2009dd 		& 0.46 & 1.1 & 0.83  & 12 	& 7.3 	& 5.5  & 130 & 2.1 & UBVRI 	 &   2\\
2010aj 	    & 0.09 & 1.3 & 1.05  & 11 	& 7.5 	& 4.3  & 94  & 0.5 & UBVRI 	 & 2\\
2013ab      & 0.47 & 1.2 & 0.71  & 13   & 8.1   & 6.8  & 116 & 2.6 & uvo & 8\\
2013ej		& 0.16 & 1.4 & 1.27	 & 19	& 8.8	& 6	   & 112 & 1.1 & uvo & 9
\enddata
 \tablenotetext{}{1. \cite{Clocchiatti96,DallOra14} 2. \cite{Inserra13}
3. \cite{Pastorello04} 4. \cite{Spiro14} 5. \cite{Hendry06,Maguire10} 6. \cite{Roy11} 7. \cite{Inserra12} 8. \cite{Bose15a} 9. \cite{Bose15b}} 
\tablenotetext{$^\dagger$}{The \Ni\ mass used to calculate $\eN$ and $\Le$. It is derived based on comparison of Eq. \ref{eq:QNi} to the observed optical luminosity, without additional bolometric corrections. It is therefore lower than the true \Ni\ mass in these SNe. Yet, it is the more accurate value to use when calculating $\eN$ and $\Le$ in SNe where only optical (and possibly UV) luminosity is available (see text).} 
\tablenotetext{$^{\dagger\dagger}$}{Luminosity is integrated over the observed bands without additional bolometric corrections. }
\tablenotetext{a}{Sparse data. A gap in the data between 65 d (on the plateau) and 113 d (during the sharp fall near the end of the photospheric phase). Luminosity at 75 d is estimated by extrapolation of the data at 56 d and 65 d.}
\tablenotetext{b}{Sparse data. A gap between 25 d and 66 d (both on the plateau) and another gap between 66 d and 98 d (at the end of the sharp fall or already on the \Ni\ tail). Luminosity at 50 d is estimated by interpolation of the data at 25 d and 66 d while luminosity at 75 d is estimated by extrapolation of these two data points. $\tN$ is taken as 98d.}
\tablenotetext{c}{Sparse data. First data point at 29 d. Luminosity at 25 d is estimated to be the same as at 29 d. The transition time to the \Ni\ tail is not observed. It is between 116 d (last data point on the plateau) and 153d (first data point in the tail). Here we take $\tN=130$ d.}.
\label{table:OptSample}
\end{deluxetable*}

\section{The sample}\label{sec:sample}
We compiled from the literature 24 well observed type II SNe for which detailed bolometric or pseudo-bolometric\footnote{We refer to light curve as bolometric if it includes bolometric corrections (e.g., via blackbody fits) that account for light that falls out of the observed bands. Pseudo-bolometric light curves account only for the light seen in the observed bands.} light curves are available. The light curves are taken from the references in tables \ref{table:BolSample} and \ref{table:OptSample} and are based on the distances and extinctions used in each reference. An exception is SN\,2013by for which we computed the bolometric light curve ourselves based on blackbody fits to the photometry from \citet{Valenti15} assuming no significant extinction and taking a distance modulo $\mu=30.84$. For a SN to be included in our sample, we required a reasonable coverage of the photospheric phase, namely first detection less than 30\,d after the estimated time of explosion, and several data points along the photospheric phase that enable a reasonable constraint on the luminosity evolution during that phase. We further required at least two measurements along the \Ni\ tail so that the mass of \Ni\ could be constrained. 

The analysis in this paper is sensitive to the light curve starting about 20 d after the explosion (we verify that below) and up to the \Ni\ tail. At these times most of the observed light is in the optical and IR. All the light curves we compile include the light emitted in the optical bands but some are missing the IR light. We therefore divide the 24 SNe to two samples. The first is composed of 13 SNe for which the published light curves provide a good approximation to the true bolometric light at the time of interest. These are light curves that include bolometric correction or pseudo-bolometric light curves that are base on UV/Opt/IR  or Opt/IR light. We refer to this sample as the bolometric sample. The SNe in this sample and their properties are given in table   
\ref{table:BolSample}. The second sample has 11 SNe where IR luminosity is missing, and thus their light curves may miss a significant fraction of the emitted luminosity. We refer to this sample as the optical sample, and its SNe are listed in table   
\ref{table:OptSample}. As described below we first derive the observables of the bolometric sample and then, based on the results of the bolometeric sample, we derive properties of the optical sample that can be used for our analysis.

For each SN in the bolometric sample we estimated first the \Ni\ mass based on the luminosity during the tail, and on equation \ref{eq:QNi}, assuming that $Q(t>\tN)=\Lb(t>\tN)$. In all but two SNe, the \Ni\ tail evolved as expected, namely a decay at a rate consistent with 0.98 mag/100\,d. The light curves of SNe 2005cs and 2007od do not follow the expected evolution. We cautiously kept them in the sample, using the observed tail to get a rough estimate of the \Ni\ mass for SN\,2005cs and a value taken from the literature for SN 2007od \citep{Inserra11}, where dust may strongly affect the observed tail luminosity. The \Ni\ masses that we find are given in table \ref{table:BolSample}. The mass found for each SNe is very similar to the value derived in the paper from which we took the SNe light curves (always within the uncertainly and up to 10-20\% from the best estimated value).

Next we determined $\tN$ for each SN. This can be done accurately (to within $\pm 10$ d) in all SNe but one (SN 2012aw), where $\tN$ was taken between the last data point of the photospheric phase and the first data point on the \Ni\ tail (see table \ref{table:BolSample} for details). For each SN we found the luminosities at days 25, 50, and 75 denoted as $L_{25}$, $L_{50}$ and $L_{75}$ respectively, by linear interpolation (in $t-log[L]$ space). As a measure of the decline rate we define $\dm \equiv -2.5\log_{10}(L_{75}/L_{25})$. We then calculate $\eN$, $\Le$ and ET for each SN. 

For all SNe with available spectra we measure the ejecta velocities, and in a single case (SN 2009md) we take it from the literature (spectra for the majority of the SNe were obtained from WISEREP; \citealt{yaron12}). These velocities are used in section \ref{sec:SN_properties}. Traditionally, the Fe\,II $\lambda5169$ absorption line velocity, as measured in mid-plateau, is considered a good proxy for the velocity of the photosphere (e.g., \citealt{Schmutz90}; \citealt{Dessart05}). Briefly, as often done before \citep{Poznanski09,Poznanski10,Poznanski15}, we cross correlate the spectra with a library of high signal-to-noise spectra for which the velocity of the $\lambda5169$ line has been measured directly. The velocity from the cross-correlation and its uncertainty are then propagated to day 50 past explosion, following \citet{Nugent06}, who showed that photospheric velocities of SNe II-P follow a tight power law relation. We use the improved determination of the phase dependance of the velocity by \citet{Faran14a}. We note that \citet{Faran14} found that SNe II-L have a different, slower, velocity evolution, with some scatter. As a result, our calculations may slightly underestimate the velocity for such SNe.

We can use the bolometric sample to evaluate the accuracy of observables that are derived based on light curves that lack IR luminosity. We repeat the derivation described above for three of the SNe in the bolometric sample, where UBVRI (or BVRI) and bolometric light curve are provided in the same paper (i.e., using the same data, estimated distance and extinction). The derivation of all the observables is done as if the optical light curve is bolometric. Namely, $M_{Ni}$ is derived by comparison of the optical light in the \Ni\ tail to equation \ref{eq:QNi}, while $\eN$ and $\Le$ are calculated by time integration of the optical light. Comparison of the quantities derived based on optical light alone to those derived based on bolometric light show that in all observables that are linear with the luminosity (i.e., $L_{25,50,75}$, $M_{Ni}$ and ET) the usage of optical light only leads to an underestimate by about a factor of 2, and in some cases even by a factor of 3. However, the difference in the dimensionless observables is much smaller -- a moderate factor of $\approx 1.2--1.4$ in the measurement of $\eN$, a difference of 0.01--0.08 in $2.5\log (\Le)$ and a difference of $\approx 0.1$ mag/50\,d in $\dm$. The reason that these dimensionless quantities are rather accurate also when only optical light is considered is that they are basically a ratio of the luminosities observed at different times and therefore they are insensitive to the absolute value of the missing bolometric correction factor. Instead they are only sensitive to the time evolution of the correction factor, which turns out to be not too large. Based on these results we restrict the usage the optical sample in our analysis to dimensionless quantities. These are given in table \ref{table:OptSample}. We also provide, for completeness, the value of the \Ni\ mass we use for the derivation of $\eN$ and $\Le$, although these are not the true \Ni\ masses in these SNe as they are derived using the optical light curves as if they were bolometric. 
 
Finally, in order to test the effect of missing data at early time (in about half of the sample the first data point is around day 20), as well as the effect of missing UV light, we used 8 SNe in our two samples, for which data starts no more than 10\,d after the estimated time of explosion. We compared $\eN$ and ET (the only two quantities affected by early data) of these SNe when measured with and without the data before day 20. The typical difference in $\eN$ is $<10\%$ and in ET it is $<20\%$. We therefore conclude that the effect of missing early data and/or UV light is not a major source of uncertainty.
This result also implies that the uncertainty in the estimated time of explosion (which is typically better than 10 days) also does not introduce a significant uncertainty to the quantities we measure.

\section{ \Ni\ contribution in observed type II SNe}\label{sec:NiContribution}
The values of $\eN$ in our sample are all, except one, in the range 0.09--0.71, with most SNe having $\eN=0.3-0.6$ (tables \ref{table:BolSample} and \ref{table:OptSample}). The exception is SN 2009ib with $\eN=2.6$. The fraction of the time weighted integrated luminosity that is contributed by \Ni\ is $\frac{\int^{\tN}_0 t\,\Q\,dt}{\int^{\tN}_0 t\,\Lb\,dt}=\frac{\eN}{1+\eN}$. Thus, the observed values of $\eN$ in our sample  (0.09--2.6) imply that the range of \Ni\ contribution to the time weighted integrated luminosity is 8\%-72\% with a typical value of 30\%.  
These values indicate that \Ni\ has a non-negligible contribution to the photospheric emission for most of the sample. But what is this effect on the observed light curve and can it be quantified?
 
$\eN$ and $\Le$ have the advantage of being observables that are independent of radiative transfer. But, for that they must be integrals and as such they do not hold information on the exact shape of the light curve. Namely, knowing their values does not enable us to determine exactly how the light curve would have been looked like if there were no \Ni. The reason is that in order to remove the \Ni\ contribution we need to know the exact \Ni\ distribution and to calculate its effect on the radiative transfer. Yet, $\eN$ and $\Le$ can provide quantitative  constraints on  the  effect of  \Ni\ on the observed light curve.  

\begin{figure}
\epsscale{1.1}
\plotone{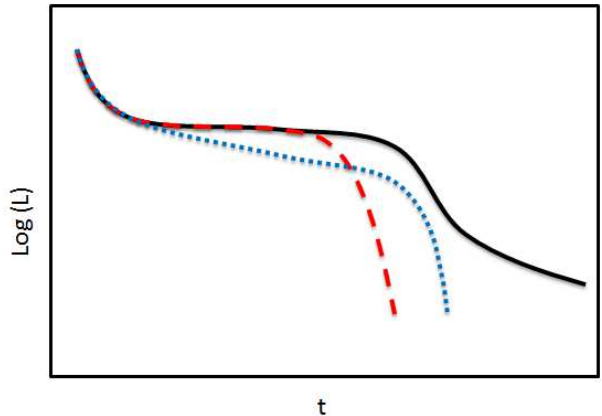}
\caption{A schematic illustration of the two effects that \Ni\ can have on the light curve for the same value of the contribution fraction $\eN$. The black solid line is the observed bolometric light curve. The red dashed line is a sketch of the cooling envelope emission if the main effect of \Ni\ is to prolong the plateau.  This scenario is is more likely if \Ni\ is concentrated in the core. The blue dotted line is a sketch of the cooling envelope emission if the main effect of the \Ni\ is to flatten the slope of the plateau. This effect is expected to dominate if \Ni\ is mixed through the envelope. The actual effect is most likely a combination of both.}
\label{fig:Le_sketch}
\end{figure}

To obtain that, we use the expectation that \Ni\ is produced in the core and that its fraction drops, or at most remains constant, with radius. This implies that the relative contribution of \Ni\ increases with time, reaching a peak near the end of the photospheric phase.  Thus, the effect of \Ni\ on the light curve can be either flattening (i.e., reducing the decline rate) or extending the duration of the photospheric phase (or a combination of both effects). In  Fig.
(\ref{fig:Le_sketch}) we sketch the two extreme possibilities of \Ni\ effect on the light curve. The  first  is  that \Ni\ is  extending the plateau  duration without
affecting the  decline rate.  In this case  the cooling envelope  contribution underlying  the combined
`observed'  light curve  would have  looked like  the dashed  red line  in Fig.  (\ref{fig:Le_sketch}).
Alternatively, \Ni\  could make the last stages  of the plateau  brighter (i.e., making it  flatter) without
significantly affecting the plateau duration, and the  cooling envelope emission would look like the dotted blue line in Fig. (\ref{fig:Le_sketch}).  The true effect is  a combination  of the two, but what
is the  actual possible effect in  each scenario and  which one is more  likely to be the  dominant? To
address  this question  we consider  each  scenario in  some  detail and  estimate (quantitatively)  its
possible effect on the SNe  in our sample.

First, consider the possibility that \Ni\ only extends the plateau without significantly affecting its decline rate. Namely, \Ni\ does not affect the light curve up to the time that $L_e$ starts fading, at which point it becomes the main power source, thereby extending the plateau. This behavior is expected to dominate   when \Ni\ is concentrated in the core and is not well mixed into the envelope. Then, the luminosity generated by \Ni\ power diffuses to the observer only after almost all the shock deposited energy has already been leaked out of the envelope\footnote{ An example of this scenario, as well as a discussion on the effect on the plateau duration, is given in \citet{Kasen09}. Their figure 2 presents numerical modeling of the light curve of a supernova with varying amounts of \Ni\ (including no \Ni). Since the \Ni\ in their modeling is concentrated towards the center of the ejecta, the light curve of the different cases deviates significantly only at late times, and \Ni\ is found to significantly extend the plateau. Another example can be seen in figure 12 of \citet{Bersten11}, that compares numerical modeling of three cases: without \Ni, with \Ni\ concentrated near the center, and with \Ni\ evenly mixed throughout the envelope. The first and the second cases show a similar light curve up to the point where $L_e$ fades and the \Ni\ power becomes the dominant energy source.}. 
If the plateau is flat (i.e., $\Lb$ is constant) then the relation between $\eN$ and the plateau extension in this scenario is analytic. Denoting $t_p$ and $t_{p,e}$ as the durations of the plateaus of $\Lb$ and $L_e$, respectively, then in this scenario $\frac{\int^\tN_0t\,\Lb\,dt}{\int^\tN_0 t\,L_e\,dt}=\frac{t_p^2}{t_{p,e}^2}$, where from Eqs. (\ref{eq:ET2}) and (\ref{eq:etaNi}) we find $\frac{\int^\tN_0t\,\Lb\,dt}{\int^\tN_0 t\,L_e\,dt}=(1+\eN)$ . Thus, in this case $t_p =t_{p,e}(1+\eN)^{1/2}=(1.15-1.25)t_{p,e}$, where the final equality is for $\eN=0.3-0.6$.  Assuming instead $\Lb$ with a constant decline rate of 1 [2] mag/100d and $t_p \approx 100$ d, the observed range $\eN=0.3-0.6$ corresponds to  $t_p =(1.2-1.35)t_{p,e}$ [$t_p =(1.25-1.5)t_{p,e}$].


\begin{figure}
\epsscale{1.1}
\plotone{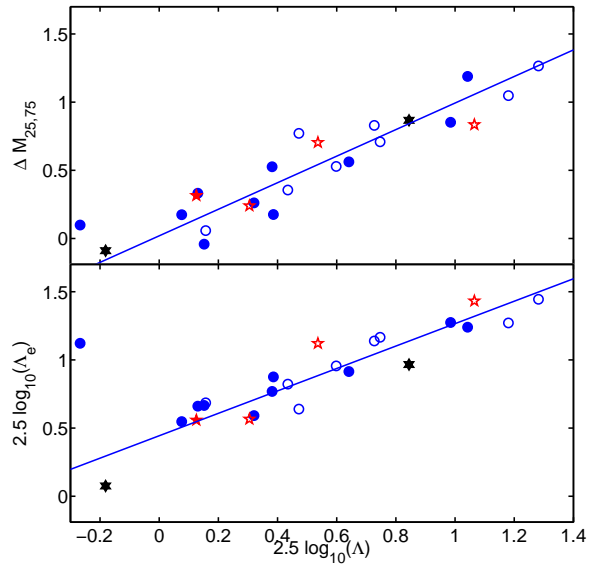}
\caption{$2.5\,\log_{10}(\Lambda)$ as a function of $\dm$ ({\it top}) and $2.5\,\log_{10}(\Le)$ ({\it bottom}). Blue circles are SNe with detailed light curves and tight constrains on $M_{Ni}$. Red pentagrams are SNe with sparse light curves (SNe 1992H, 1995ad, 2003z and 2012aw), while  black hexagrams are SNe with large uncertainty in $M_{Ni}$ (SNe 2005cs and 2007od). Filled symbols are SNe from the bolometric sample (Table~\ref{table:BolSample}). Empty symbols are SNe from the optical sample  (Table~\ref{table:OptSample}). The solid lines are the best linear fits. In the top panel it is $\dm=2.5\,\log_{10}(\Lambda)$ and in the bottom panel it is $2.5\,\log_{10}(\Le)=0.44 + 0.82 \cdot 2.5\,\log_{10}(\Lambda)$.}
\label{fig:LambdaVSdm}
\end{figure}

Next, we consider the other extreme possibility, namely  that the plateau duration is not affected by
\Ni\ (or $t_p=t_{p,e}$) and that instead the \Ni\  changes only the luminosity along the plateau. Since
\Ni\ is  always negligible at  early times the  effect must  be an increase  in the luminosity  at late
times, namely  more \Ni\ results in  a flatter (or  even rising) plateau.  This is the effect  which is
expected to dominate  if \Ni\ is well  mixed throughout a significant fraction of the envelope. Efficiently  mixed \Ni\ generates
luminosity that diffuses  to the observer at  earlier stages, simultaneously with  the cooling envelope
emission. Without the cooling envelope emission, the  \Ni\ generated luminosity would have looked in this case like a  broad peak (similar to that observed in SN 1987A). The main effect  on the  plateau duration  is then via  increase in  the envelope
opacity.    The flattening effect on \Ni\ is seen clearly in  the numerical
light curve of  the fully mixed case in figure 12 of \cite{Bersten11}. 

The change in  the decline rate
that \Ni\  induces in this case can be estimated using $\Le$,  which is a ratio between the  luminosity at early
time (day 25) and the time-weighted average luminosity. Thus, if the duration of the plateau is roughly
constant in all SNe  then we expect $\Le$ to be  a measure of the decline rate of $L_e$. To  verify this, and to
calibrate  $\Le$, we  define its  bolometric  luminosity counterpart:  
\begin{equation} 
	\Lambda  \equiv \frac{L_{25}  \cdot(80\,d)^2}{\int^\tN_0t\,\Lb\,dt}. 
\end{equation}  
Fig. (\ref{fig:LambdaVSdm})  shows
the tight  linear correlation  between $\dm$  and $2.5\,\log_{10}(\Lambda)$,  which is  consistent with
$\dm=2.5\,\log_{10}(\Lambda)$ (the  constant $(80\,d)^2$ in the  definition of $\Lambda$ and  $\Le$ was
chosen to obtain  this equality). Thus, $2.5\,\log_{10}(\Lambda)$  is a good estimator  of the observed
decline rate (in units of mag/50\,d) and if the \Ni\ does not affect the plateau duration then $\Le$ is
a        good        estimator        of         the        decline        rate        of        $L_e$.

Fig. (\ref{fig:LambdaVSdm}) also shows $2.5\,\log_{10}(\Le)$ as a function of $2.5\,\log_{10}(\Lambda)$. 
The measured values are narrowly distributed around $2.5\,\log_{10}(\Le)=0.44 + 0.82 \cdot 2.5\,\log_{10}(\Lambda)$ (with the exception of SN 2009ib). This implies two interesting points. First, in this scenario the \Ni\ reduces the decline rate in flat SNe (with small $\Lambda$) by about 1 mag/100 d, while SNe with fast decline (large $\Lambda$) are less affected. For the majority of the SNe in our sample the flattening in this scenario is in the range of 0.5-1 mag/100 d.
Second, the tight correlation between $\Lambda$ and $\Lambda_e$ indicates that while  \Ni\ affects the observed decline rate, it is not the only source for the difference between SNe in the observed decline rates. Namely, different observed decline rates often reflects different decline rates of the cooling envelope emission. 
 
The two scenarios discussed above provide bounds to the effect of \Ni\ on the light curve. In reality it is a combination of both (i.e., both a flattening and an extension of the plateau), where the relative importance depends on the level of \Ni\ mixing in the envelope. Yet, there are several lines of evidence that support that  flattening is significant, and possibly the dominant effect on the light curve:

(i) Flattening is expected if \Ni\ is well mixed through the envelope. The best clue regarding \Ni\ mixing in SN explosions of progenitors with massive hydrogen-rich envelopes is SN 1987A and similar events \citep[e.g.][]{Kleiser11}, which have progenitors with similar masses and composition but smaller radii than those of regular type II SNe. The handful of SNe of this type that were well observed are all showing rather similar light curves, which often look like a clone of SN 1987A up to a normalization factor \citep[][ and references therein]{Pastorello12}. Detailed light curve modeling of SN 1987A favor an efficient \Ni\ mixing throughout the hydrogen envelope \citep[e.g., ][]{Shigeyama90,Blinnikov00,Utrobin14}. 

(ii) The first scenario -- no \Ni\ mixing and a significant extension of the plateau -- requires a fine tuning of the amount of \Ni. The reason is that the photospheric phase can be separated in this case into two phases -- first a cooling envelope phase where \Ni\ is negligible and later a \Ni\ dominated phase. The luminosity of the two phases depends on different explosion parameters (the first phase is determined mostly by the progenitor radius and to some extent by the explosion energy and progenitor mass, while the second phase depends only on the amount of \Ni\ mass). Therefore, without finely tuning the amount of \Ni, a transition between the two phases is expected to be observed during late stages of the plateau, either as a bump if there is too much \Ni\ or a dip if there is not enough. This can be seen in \cite{Kasen09} (their Fig. 2) where $M_{Ni}$ is varied by a factor of 4 (and so does $\eN$ between 0.5 and 2). This variation is enough to change the light curve from a dip near the end of the plateau to a bump. In our sample the plateaus do not show any transition between two phases, even though the \Ni\ contribution varies significantly ($\eN=0.09-2.6$) including a case where \Ni\ dominates the photospheric light curve. 

(iii) In the second scenario, where \Ni\ is well mixed, its contribution becomes significant earlier, near the middle of the plateau (if SN1987A is a reasonable example for \Ni\ contribution then around day 50), and from that point, both \Ni\ and cooling envelope are significant until the plateau ends. Thus, instead of a transition between the two phases, a flattening is expected to be seen around the middle of the plateau. Such a flattening is indeed observed in many SN light curves (e.g., \citealt{Anderson14}; the transition from s1 to s2 in their notation).

(iv) Each scenario predicts different correlations between observables. If \Ni\ results in flattening of the light curve then an anti-correlation is expected between the importance of \Ni\ contribution ($\eN$) and the decline rate. On the other hand, if \Ni\ extends the plateau no such anti-correlation is expected and instead a correlation between $\eN$ and $\tN$ is predicted. As we show in the next section we find a highly significant anti-correlation between $\eN$ and $\dm$ and at most a weak correlation between $\eN$ and $\tN$.

These arguments are all suggesting that $2.5\,\log_{10}(\Le)$ is not only an upper limit on the decline rate of the light curve without \Ni, $L_e$, but that it also provides a rough estimate of this decline rate. 

{ To conclude, \Ni\ contribution to the photospheric emission is significant in type II SNe, although in most it is not the main source of the observed luminosity. Typically it contributes around 30\% of the time weighted integrated luminosity, but in some SNe it contributes less than 10\% and in the extreme case of     
SN2009ib it contribute 72\%. This contribution flattens and extends the plateau. In typical SNe, the maximal possible extension of the plateau is by about 25\% while the maximal possible flattening is by about 1 mag/100 d.} The actual effect is a combination of both, however, several line of evidence suggest that flattening is significant. Finally, while \Ni\ most likely affect the decline rate of the plateau, the observed range of decline rates is clearly dominated by the decline rates of the cooling envelope emission and cannot be a result of \Ni\ alone.

\section{Correlations between observables}\label{sec:correlations}

\begin{deluxetable*}{p{1cm}ccccccccccccc}
\tabletypesize{\scriptsize}
\tablecolumns{12} 
\tablewidth{0pt}
\setlength{\tabcolsep}{0.0in}
 \tablecaption{Correlations between various observables}
 \tablehead{
  & $\eN$ & $\Le\,^a$ & $\dm\,^a$ & $\tN\,^b$ & $L_{25}\,^c$ & $L_{50}\,^c$ & $L_{75}\,^c$ & $M_{Ni}\,^c$& $v_{50}\,^d$& ET$\,^c$ &  $ET \cdot v_{50}\,^d$ & $ET/v_{50}\,^d$\\&&&&&&&&&& $(\propto \sqrt{EM}R)$ & $(\propto ER)$ & $(\propto MR)$ }
\startdata
$\eN$   &---& N/A 	&$(-)0.003$  & 0.05 &    N/A 		& N/A  & N/A & 0.02& N/A & N/A 	& N/A 		& N/A\\
$\Le$   &   & ---	&$<10^{-4}$&$(-)$0.04 &    0.03 	&0.06  & N/A & N/A & N/A & N/A 	& N/A 		& N/A\\
$\dm$   &   &		&	---	   &$(-)$0.04 &$3\cdot 10^{-4}$&0.002 &0.06 &N/A& 0.06 &0.005& 0.02	   	& 0.04\\ 
$\tN$	&	&		&		   & ---  &    N/A 		& N/A  & N/A &0.07 & N/A & N/A		& N/A	   	& N/A\\	
$L_{25}$&	&		&		   &   	  &    --- 		&	   &	 &0.06 &$3\cdot 10^{-4}$&$<10^{-4}$&$<10^{-4}$	&$<10^{-4}$\\	 	
$L_{50}$&	&		&		   &  	  &     		& ---  &	 &0.04 &$ 10^{-3}$&$<10^{-4}$&$<10^{-4}$	&$<10^{-4}$\\		 
$L_{75}$&	&		&		   &	  &     		&	   & --- &0.008 &$2\cdot 10^{-3}$&$<10^{-4}$&$<10^{-4}$	&$<10^{-4}$\\		 
$M_{Ni}$&	&		&		   & 	  &     		&	   & 	 & --- & N/A & 0.03		& 0.01	   	& 0.05	\\	 
\tablenotetext{}{The significance of correlations between various observables derived in this paper. Minus sign, `$(-)$', in the table marks a significant anti-correlation.The values in the table are the probability that there is no ranked correlation (or anti-correlation) as obtained by Monte Carlo simulations in the following way. In each simulation we obtained $10^5$ realizations, each constructed of random pairings of the values in our sample (with no repetitions), and calculate the Spearman's rank correlation coefficient of each realization. The [anti]correlation significance is defined as the fraction of realizations that have a  coefficient that is larger [smaller] than that of the actual observed sample. 
N/A represents very low correlation significance (probability of no correlation $>0.1$).}
\tablenotetext{a}{Correlations are calculated based on all the SNe in our two samples (24 SNe)}
\tablenotetext{b}{Correlations are calculated based on all the SNe in the two samples with light curves that are sampled well enough and $\tN$ is measured accurately (20 SNe; SNe 1992H, 1995ad, 2003Z and 2012aw have sparse data and are therefore excluded.)}
\tablenotetext{c}{The correlations are calculated based only on the bolometric sample (all 13 SNe in Table \ref{table:BolSample}). SN 2012aw is excluded in the correlation with $\tN$ due to sparse data.  }
\tablenotetext{d}{The correlations are calculated based only on SNe in the bolometric sample with measured $v_{50}$ (11 SNe with measured $v_{50}$ in Table \ref{table:BolSample}). SN 2012aw is excluded in the correlation with $\tN$ due to sparse data.}
\label{table:Correlations} 
\end{deluxetable*} 

Table \ref{table:Correlations} list the significance of correlations (and anti-correlations) between every two observables listed in tables \ref{table:BolSample} and \ref{table:OptSample}. Here we focus on the correlations of the two new dimensionless observables that we introduce, $\eN$ and $\Le$, which are relevant for the effect on \Ni\ on the light curve.

\begin{figure}[!t]
\epsscale{1.1}
\plotone{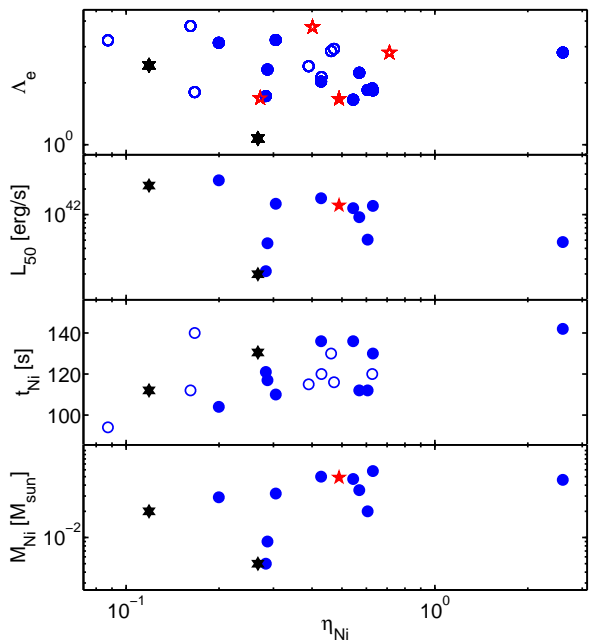}
\caption{A scatter plot of the measure of \Ni\ importance, $\eN$, as a function of $\Le$, $L_{50}$, $t_{Ni}$ and $M_{Ni}$. Symbols are the same as in Fig. (\ref{fig:LambdaVSdm}). In each panel only SNe used to calculate the correlation are included (see table \ref{table:Correlations}). $\eN$ shows no correlation with  $\Le$, and $L_{50}$ while a marginally significant correlation is found with $t_{Ni}$ and $M_{Ni}$.} 
\label{fig:etaNivsX}
\end{figure}

$\eN$ does not show a significant correlation to most of the observables. Specifically, it seems to be uncorrelated with $\Le$, which is a measure of the cooling envelope shape, and with the plateau luminosity  $L_{25}$, $L_{50}$ and $L_{75}$ (see illustration in figure \ref{fig:etaNivsX}). It also shows no correlation with $ET$, $ET \cdot v_{50}$ and $ET/v_{50}$ (see discussion in section \ref{sec:SN_properties}). A marginally significant correlation (probability of no correlation $0.05$) is seen with the plateau duration and a slightly more significant correlation (probability of no correlation $0.02$) with $M_{Ni}$ (see figure \ref{fig:etaNivsX}). 

\begin{figure}[!t]
\epsscale{1.1}
\plotone{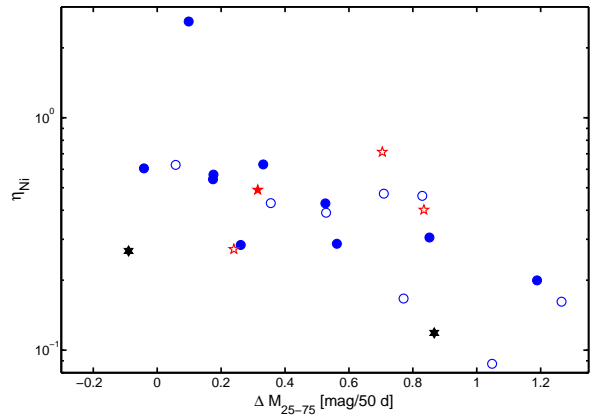}
\caption{The measure of \Ni\ importance, $\eN$, as a function of the decline rate between days 25 and 75, $\dm$. Symbols are the same as in Fig. (\ref{fig:LambdaVSdm}). A clear anti-correlation, with a large scatter, is observed between $\eN$ and $\dm$, indicating that \Ni\ importance decreases the faster is the decline.} 
\label{fig:dmVSetaNi}
\end{figure}

The only highly significant finding (probability of no correlation $0.003$) is an anti-correlation with the bolometric decline rate between days 25 and 75, $\dm$. Fig. (\ref{fig:dmVSetaNi}) shows a scatter plot of $\eN$ vs. $\dm$. A roughly linear anti-correlation is clearly seen. Namely, the role of \Ni\ during the photospheric phase is more prominent in SNe with a smaller decline rate (a `flatter' plateau). The scatter around the correlation is considerable and it is most likely a combination of intrinsic scatter and the inhomogeneity of our sample.  

\begin{figure}[!t]
\epsscale{1.1}
\plotone{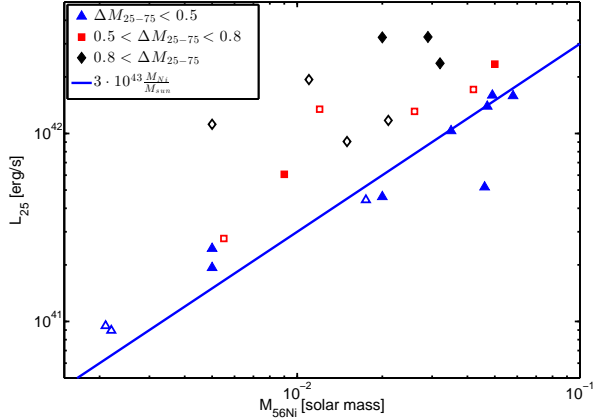}
\caption{Bolometric luminosity at day 25 as a function of \Ni\ mass. SNe are marked according to their decline rates (see legend). The blue line shows a linear relation between $L_{25}$ and $M_{Ni}$, normalized to fit the slowest declining SNe (The slow declining SN that is significantly below the lin eis SN 2009ib). The figure shows a strong luminosity-$M_{Ni}$ linear correlation, where faster declining SNe are on average brighter than slow declining SNe with the same $M_{Ni}$.}
\label{fig:L25VSMNi} 
\end{figure}

\begin{figure}[!t]
\epsscale{1.1}
\plotone{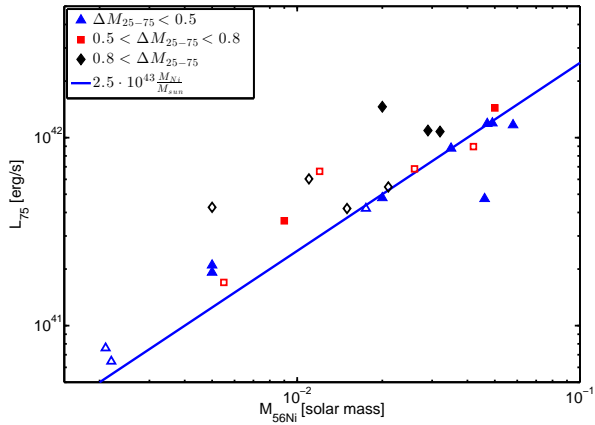}
\caption{Bolometric luminosity at day 75 as a function of \Ni\ mass. SNe are marked according to their decline rates (see legend). Also marked is SN 1987A. The blue line shows a linear relation between $L_{25}$ and $M_{Ni}$, normalized to fit the slowest declining SNe.}
\label{fig:L75VSMNi}
\end{figure}

The anti-correlation between \Ni\ contribution and the decline rates can also be seen when the well known correlation between the plateau luminosity and the \Ni\ mass is plotted. Fig. (\ref{fig:L25VSMNi}) shows the luminosity at 25 days after explosion, $L_{25}$, as a function of the \Ni\ mass. An approximate linear relation is seen, similarly to the finding of previous works \citep[e.g.,][]{Hamuy03}. In order to show the relation to the \Ni\ contribution and decline rate, we use different symbols for SNe with different values of $\dm$. SNe that decline fast are significantly above those that decline slowly. Namely, for the same amount of \Ni\, fast declining SNe are brighter and thus have a smaller $\eN$. This is consistent with the results of \cite{Valenti15}. 
Fig. (\ref{fig:L75VSMNi}) is similar to Fig. (\ref{fig:L25VSMNi}), but with $L_{75}$ instead of $L_{25}$. While fast declining SNe are still brighter on average at day 75 (for the same amount of \Ni) , the difference from slow declining SNe is reduced, as expected. 

We have seen in section \ref{sec:NiContribution} that for SNe with a high value of $\eN$, \Ni\ must have an observable effect either via flattening the plateau or by extending it or, most likely, both.
The significant anti-correlation between $\eN$ and $\dm$ supports a picture were flattening is a major effect. This in turn requires \Ni\ to be mixed into a significant part of the envelope. The marginal correlation with $\tN$ suggest that \Ni\ also extends the plateau, although not significantly. 

\begin{figure}[!t]
\epsscale{1.1}
\plotone{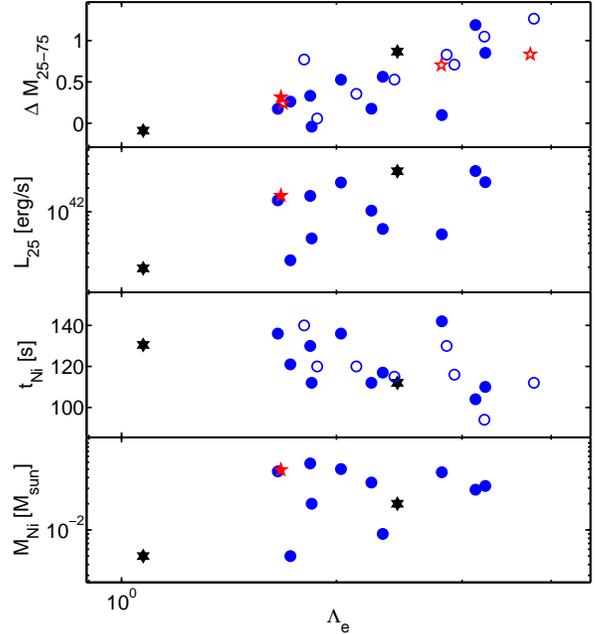}
\caption{A scatter plot of $\Le$, a probe the shape of the light curve that we would have seen if there were no \Ni, as a function of $\dm$, $L_{25}$, $t_{Ni}$ and $M_{Ni}$. Symbols are the same as in Fig. (\ref{fig:LambdaVSdm}). In each panel only SNe used to calculate the correlation are included (see table \ref{table:Correlations}). $\Le$ shows a very strong correlation with  $\dm$, a marginally significant correlation with $L_{25}$ and marginally significant anti-correlation with $t_{Ni}$. $\Le$ shows no correlation with $M_{Ni}$.} 
\label{fig:Lambda_eVSx}
\end{figure}

Next we examine the correlations of $\Le$, some of which are illustrated in figure \ref{fig:Lambda_eVSx}. $\Le$ probes the shape of $L_e$, the light curve that we would have seen if there were no \Ni. It is larger when $L_e$ declines faster and/or when its duration is shorter. $\Le$ shows a very strong correlation with $\dm$. This indicates that the observed decline rate is dominated by the cooling envelope emission. Namely, while \Ni\ most likely affect the decline rate it is not the source of the range of decline rates that are observed. Instead this range is dominated by the cooling envelope emission.  $\Le$ also shows a marginal anti-correlation with $\tN$ suggesting that the variance that we see in plateau durations is at least partially determined by $L_e$. Finally $\Le$ is also correlated with the plateau luminosity. This correlation is most likely related to the correlation between the decline rate and the luminosity \citep{Anderson14}, whose origin is still unknown. Other than that  $\Le$ shows no correlation to any other observable we examined.  Interestingly, this includes $M_{Ni}$ and $\eN$, suggesting that the shape of $L_e$ is not correlated with the \Ni\ amount in the ejecta or its importance in shaping the observed light.

\section{Constraining the progenitor and explosion properties using ET and $v_{50}$}\label{sec:SN_properties}

As discussed in \S\ref{sec:theory} the total contribution of the cooling envelope to the (time-weighted) energy release, ET, can be directly obtained from  observation without major uncertainties related to the details of radiative transfer. This observable is directly related to the explosion energy and progenitor structure. This relation was explored recently by \citet{Shussman16}. Here we use their results to study our bolometric sample (Table~\ref{table:BolSample}), for which we obtained a reliable measure of ET.

\citet{Shussman16} used a large set of numerically calculated red supergiants to study what ET can teach us on the progenitor and explosion energy. In general $ET \propto \sqrt{EM}R \propto MRv$ where $E$, $M$, $R$ and $v$ are energy, mass, radius and velocity that characterize the explosion and the progenitor. Specifically they find three approximations, each for a different characteristic E, M and R:
\begin{eqnarray}\label{eq:ET}
	ET &\approx& 0.42~\Eenv^{1/2}\Renv \Menv^{1/2} \nonumber \\
	   &\approx& 0.18 \Eenv^{1/2}R_*\Menv^{1/2} \approx  0.15 \Eexp^{1/2}R_*\Mej^{1/2} 
\end{eqnarray}
where $\Eexp$ is the total explosion energy carried by the entire ejecta to infinity, $\Eenv$ is the energy carried by the envelope to infinity, $\Mej$ [$\Menv$] is the ejecta [envelope] mass, $R_*$ the progenitor radius and $\Renv=\frac{\int r dm}{\Menv}$
is the mass weighted average radius of the envelope ($dm$ is mass element of the progenitor at radius $r$). The quality of the approximations vary between the three. The reasons are, first, that the cooling envelope emission depends only on the envelope properties, and therefore approximations that use only envelope characteristics, the first two in equation \ref{eq:ET}, are more accurate than the third one, which use global properties of the ejecta. Second, $ET$ also depends on the internal structure (density profile) of the progenitor, thus the first approximation which takes it into account is better than the second one. Quantitatively, the first, which depends on $\Eenv$, $\Renv$ and $\Menv$, is accurate to within 20\% for all the progenitors considered by \citet{Shussman16}, including those that lost almost all of their hydrogen envelope. The second approximation, which depends on $\Eenv$, $R_*$ and $\Menv$ is accurate to within 20\% for progenitors that retain most of their envelope, and to within a factor of 3 for progenitors that lost most of their envelope. The third approximation, which depends on $\Eexp$, $R_*$ and $\Mej$ is applicable only to progenitors where $\Menv/\Mej>0.6$, in which case it is accurate to within about 30\%.

Additional information can be extracted from ET for SNe with spectral measurements of the photospheric velocity. \citet{Shussman16} show that $v_{50}$ provides a good estimate of $\venv=\sqrt{2\Eenv/\Menv}$, and when the ejecta is dominated by the envelope it is also a good approximation of $\vej=\sqrt{2\Eexp/\Mej}$. Thus, $ET \cdot v_{50} \propto \Eenv R_* \propto \Eexp R_*$ and $ET/v_{50} \propto \Menv R_* \propto \Mej R_*$.

\begin{figure}[ht]
\epsscale{1.1}
\plotone{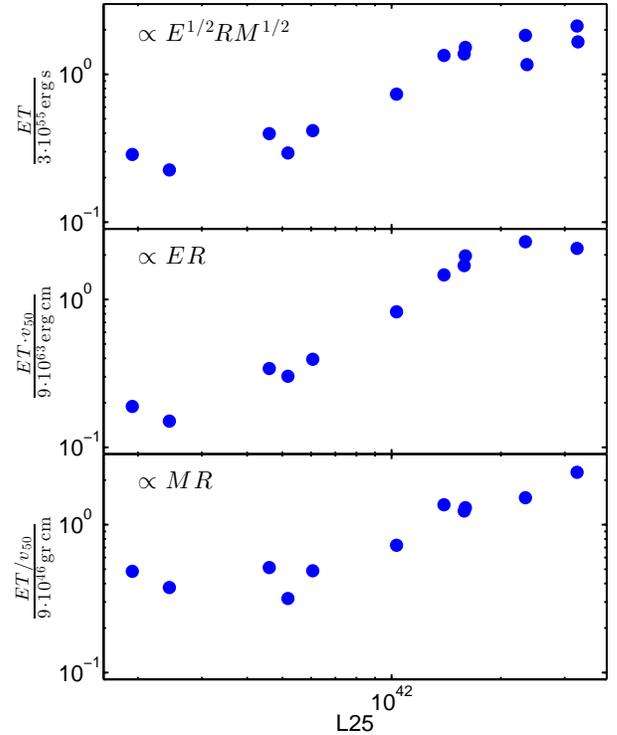}
\caption{ $ET \propto \sqrt{EM}R$, $ET \cdot v_{50} \propto ER$ and $ET/v_{50}\propto MR$ as a function of $L_{25}$. The top panel includes the 13 SNe of the bolometric sample, while the two other panels include each 11 SNe of this sample that also have a measured $v_{50}$ (Table~\ref{table:BolSample}). 
Very significant correlations are seen in all three panels, suggesting that brighter SNe are more energetic and with larger mass and/or larger radii.}
\label{fig:E0M0R0vsL25}
\end{figure}

In order to apply equation \ref{eq:ET} to observations a reliable estimate of ET is needed, which in turn requires good bolometric light curve. Therefore we use here only the 13 SNe in our bolometric sample (Table~\ref{table:BolSample}). Fig. (\ref{fig:E0M0R0vsL25}) depicts scatter plots of $L_{25}$ with $ET$, $ET \cdot v_{50}$ and $ET/v_{50}$, normalized by $0.18 (\Eenv \Menv)^{1/2}R_*$, $0.26 \Eenv R_*$ and $0.13 \Menv R_*$, respectively, where $\Eenv=10^{51}$ erg, $R_*=500R_\odot$ and $\Menv=10M_\odot$.   The logarithmic mean of all three is almost identical, 0.8, suggesting that the canonical progenitor envelope values we used for the normalization are representative for this sample. The spread in the values is largest for $ET \cdot v_{50}$, about an order of magnitude, and smallest for $ET/v_{50}$, about a factor of 3. This suggests that unless there is an anti-correlation between $\Menv$ and $R_*$ (RSGs calculated by the stellar evolution code MESA does not show such anti-correlation; \citealt{Shussman16}), the spread is dominated by variation of $\Eenv$ which in most progenitors is similar to $E_{exp}$.  This spread is consistent with the findings of \citet{Poznanski13} who found $E_{exp} \propto M_{ej}^3$ in a sample of SNe II-P with detected progenitors.

The typical values that we find are similar to those found in estimates of $E_{exp}$, $M_{ej}$ and $R_*$ via detailed hydrodynamical modeling of individual events in our sample (see the references in Table~\ref{table:BolSample}). However, \cite{Hamuy03} finds systematically larger explosion energies and ejecta masses, and lower progenitor radii. The estimates in \citet{Hamuy03} are based on general numerical models \citep{Litvinova83,Litvinova85} that solve for $E_{exp}$, $M_{ej}$ and $R_*$ based on three observables, $L_{50}$ (in V band), $v_{50}$, and the plateau duration. These models ignore the \Ni\ contribution and this can explain at least part of the discrepancy. As we find here, \Ni\ is affecting the plateau properties in most type II SNe. Our results show that \Ni\ either flatten the plateau, in which case it is most likely contributes significantly to $L_{50}$, and/or extends the plateau. Both effects, when ignored, cause overestimates of $E_{exp}$ and $M_{ej}$ and an underestimate of $R_*$.

\begin{figure}[ht]
\epsscale{1.1}
\plotone{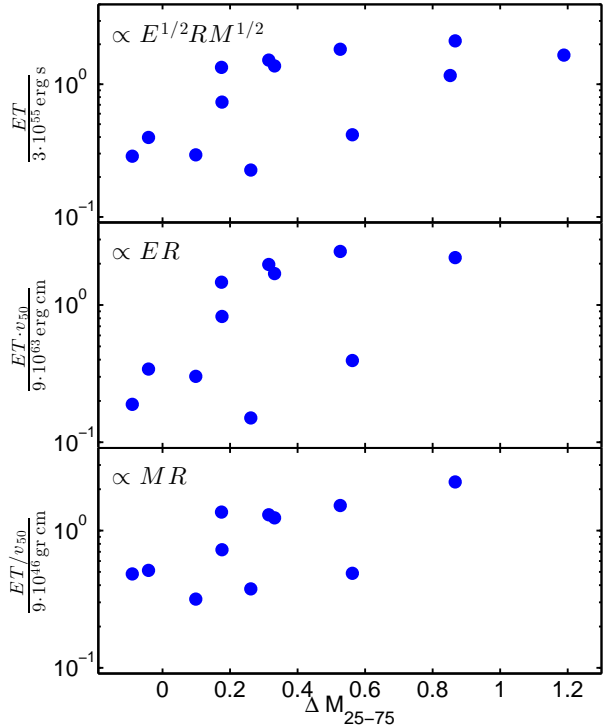}
\caption{  $ET \propto \sqrt{EM}R$, $ET \cdot v_{50} \propto ER$ and $ET/v_{50}\propto MR$ (the same samples as in Fig. \ref{fig:E0M0R0vsL25}) as a function of $\dm$. Correlations at varying levels of significance are seen in the three panels (99.5\%, 98\% and 96\% from top to bottom). This suggests that fast declining SNe are more energetic and with larger mass and/or larger radii. This is in contradiction to the suggestion that the fast decline in type II-L SNe is a result of low mass hydrogen envelope.}
\label{fig:E0M0R0vsDM}
\end{figure}

Table~\ref{table:Correlations} lists the correlation significance of $ET$, $ET \cdot v_{50}$ and $ET/v_{50}$ with other observables. No correlation is found with $\eN$ suggesting the \Ni\ relative contribution is most likely uncorrelated to the envelope properties and explosion energy. We do find however highly significant correlations with all luminosities ($L_{25}$, $L_{50}$ and $L_{75}$; See for example Fig. \ref{fig:E0M0R0vsL25}). This implies, again assuming no anti-correlation between $\Menv$ and $R_*$, that brighter SNe are more energetic and more massive and/or with larger radius. This is consistent with the estimates based on hydrodynamical modeling for individual SNe  (see the references in Table~\ref{table:BolSample}). Finally, all three observables show a significant 
correlation with $\dm$. This suggests that faster declining SNe  have larger $\Eenv$ and larger $\Menv$ and/or $R_*$. Some of the implications of this correlation are discussed in the next sub-section.

\subsection{Implications for the origin of type II-L SNe} 
The progenitor properties of type II-L SNe is still unknown. The most common suggestion is that the fast decline is a result of a small envelope mass. This possibility was first speculated by \citet{Barbon79}. Later, numerical simulations have shown that envelopes masses of $\approx 1-3 M_\odot$ can produce the observed linear declines   \cite[e.g.,][; see however \citealt{Morozova15}]{Swartz91,Blinnikov93}. Our results of positive correlation between $ET/v_{50}\propto \Menv R_*$ and $\dm$ does not support this picture.  This correlation is evident in Fig. (\ref{fig:E0M0R0vsDM}).
It shows that $\Menv R_*$ of fast declining SNe are larger by a factor of 2-3 compared to slow declining SNe. Thus, unless there is a strong anti-correlation between $\Menv$ and $R_*$, which contradicts predictions of stellar evolution models, the envelope of fast declining SNe in our sample are at least as massive as those of slow declining SNe. Moreover, for 6 out of the 11 SNe with estimate of $\Menv R_*$ there are pre-explosion images of the progenitor \citep{Smartt15} from which $R_*$ can be estimated (see table 1 in \citealt{Shussman16}). The range of radii is small, about a factor of 2, and they show no correlation with $\dm$.   
 
Another indication that faster decline is unlikely to be the result of a lower envelope mass is that if that were the case, the photospheric phase of fast declining SNe should have been much shorter. The reason is that for a given ejecta velocity the duration of the photospheric phase depends strongly on the ejecta mass (fast declining SNe have similar $v_{50}$ to slow declining ones with similar luminosity). This can bee seen for example in \cite{Swartz91} and \cite{Blinnikov93} who model type II-L SNe with low mass ($1-3\,M_\odot$) envelopes, where for all these models $\tN < 50$ d. In our observed sample, on the other hand, $\tN$ is in the range of 90--140\,d and very weakly correlates with $\dm$ if at all\footnote{ $\tN$ in the sample is narrowly clustered around the mean (120 d) where almost all SNe are in the range 110-140 d and no correlation with $\dm$ is observed. The exception are three fast declining SNe, 1995ad, 2010aj and 2013by, which have $\tN$ shorter by $\sim 10-20\%$ than the average.}. 

If, indeed, the envelope mass is not the main driver of the observed range of decline rates then what can it be? We show here that while \Ni\ most likely affects the observed decline rate its effect is only secondary to that of the intrinsic cooling envelope emission, $L_e$. The temporal evolution of $L_e$, in turn, is determined to a large extent by the pre-explosion density profile of the progenitor, we therefore speculate that the decline rate is determined mostly by the envelope structure and not by its mass.

\section{Summary}\label{sec:Summary}
We explore the effect of \Ni\ on the photospheric emission of type II SNe. We use energy conservation in a spherical outflow that expands homologously to derive a measure for the importance of \Ni\ (Eq. \ref{eq:etaNi}). This measure is obtained by time weighted integrals over the bolometric light curve and \Ni\ decay energy deposition rate. As such it is insensitive to the complicated physical processes of radiative transfer and its accuracy depends mostly on the quality of the bolometric light curve. We use a similar method to derive a measure that depends purely on the light curve shape of the cooling envelope emission that would have been seen if there were no \Ni\ in the ejecta (Eq. \ref{eq:Lambda_e}).

We compile from the literature a sample of 24 type II SNe with detailed bolometric and quasi-bolometric light curves of the photospheric phase and \Ni\ tail. The sample is heterogeneous, ranging over a factor of 30 in plateau luminosity and \Ni\ mass, and over a range of 2.5 mag/100\,d in decline rate. Namely, the sample includes luminous, intermediate and low-luminosity type II SNe as well as SNe that were classified as II-P and as II-L. We analyze this sample, calculating the new observables that we derive here and find the following results:
\begin{itemize}
\item \Ni\ contribution is significant in most of the sample. In many SNe it is the source of about 30\% of the time weighted integrated luminosity during the photospheric phase. But, there is a single SN (2009ib) in the sample where \Ni\ contributes 72\% and several SNe where it powers about 10\%.

\item \Ni\ contribution can possibly flatten and/or extend the photospheric light curve. We find that if flattening were the only effect, then SNe that  are observed to have a flat plateau would have shown a decline rate of about 1 mag/100 d in the absence of \Ni. If on the other hand the only effect of \Ni\ were to extend the plateau, then without \Ni\ the photospheric phase would have been typically shorter by a factor of about 1.15--1.35 and in most extreme cases by almost a factor of 2.

\item \Ni\ contribution cannot explain the entire range of observed decline rates, which is mostly an intrinsic property of the cooling envelope.  

\end{itemize}

Several independent lines of evidence suggest that the dominant effect of \Ni\ is flattening of the light curve. This includes a significant anti-correlation between the level of \Ni\ contribution and the decline rate (i.e., \Ni\ is more important in SNe with a flatter plateau).  This is expected if \Ni\ is mixed throughout a significant fraction of the envelope (as suggested by \citealt{Bersten11} for SN\,1999em). With this interpretation the picture that arises from our results is that flat plateaus are not a generic feature of cooling envelope emission. Instead many observed plateaus includes a significant contribution of \Ni, and without it many of the type II-P SNe would have shown a decline rate of up to 1 mag/100\,d. Note that while not being expected a-priori, there is no need for a `conspiracy' so the combination of cooling envelope and \Ni\ emission would result in plateaus. First, type II SNe exhibit a range of decline rates, thus in some cases the combined emission produces a plateau and in others it does not. Second, it seems that even without \Ni\ the cooling envelope emission often shows a moderate decline rate where the luminosity drops by less than a factor of two over 100 d. In these cases the contribution of \Ni\ is enough to make the light curve even flatter. The coincidence in type II SNe seems to be the fact that the contribution of cooling envelope and \Ni\ powered emission are comparable over a wide range of light curve luminosities. This is a result of the well known tight correlation between the luminosity and \Ni\ mass \citep{Hamuy03}, which most likely reflects a correlation between the explosion energy and \Ni\ production.

Finally, we do not find correlations between the  level of \Ni\ contribution and  most of the observed light curve properties, as well as other properties of the cooling envelope emission. On the other hand, $\Le$, a property of cooling envelope emission alone, does correlate with the plateau luminosity and decline rate.  
This suggests that the many correlations seen between properties of type II light curve (Luminosity, velocity, decline rate, etc.) seem to be intrinsic properties of the cooling envelope emission. 

We use the method to remove the effect of \Ni\ on the integrated light curve to derive an observable measure of a combination of the progenitor radius, envelope mass and explosion energy (equation \ref{eq:ET}; see \citealt{Shussman16} for more details). This measure is robust in the sense that, being a result of an integrated energy conservation equation, it is insensitive to the radiation transfer of the observed light. Applying this measure to our sample, together with a measurement of the photospheric velocity at day 50, we find, in agreement with previous studies, that brighter SNe are most likely generated by more energetic explosions with larger ejecta mass and/or progenitor radius. We also find that the decline rate is positively correlated with $M_{env}R_*$, which contradicts the popular view that fast decline in type II SNe is a result of significant envelope stripping and thus a low value of $M_{env}$. Instead it is more likely to be related to the mass distribution of the ejecta, rather than to its total mass.

\acknowledgments
We thank Iair Arcavi, Doron Kushnir, Viktoriya Morozova and
Tony Piro for useful comments. We thank Ofer Yaron for pointing out the extreme light curve of 2009ib. We thank the anonymous referee for a thorough report that helped us improving the paper. This research made use of the Weizmann interactive supernova data repository  (\texttt{www.weizmann.ac.il/astrophysics/wiserep}), as well as the NASA/IPAC Extragalactic Database (NED) which is operated by the Jet Propulsion Laboratory, California Institute of Technology, under contract with NASA. 
This research was partially supported by the I-CORE Program (1829/12). EN was was partially supported by an ERC starting grant (GRB/SN), ISF grant (1277/13) and an ISA grant. BK was partially supported by the Beracha Foundation.


\end{document}